\documentstyle[aps,epsf]{revtex}
\newcommand{\be}{\begin{eqnarray}}
\newcommand{\ee}{\end{eqnarray}}
\newcommand\del{\partial}

\begin{document}


\title{
\begin{flushright} 
{\small BNL-NT-01/11}
\end{flushright} 
One-Loop QCD Corrections to the Thermal Wilson Line Model} 
\author{J. Wirstam}

\address{Department of Physics, Brookhaven National Laboratory, Upton, NY 11973, USA. \\ email: wirstam@bnl.gov} 
\maketitle

\begin{abstract} 
We calculate the time independent four-point function in high temperature ($T$) QCD and obtain the leading 
momentum dependent terms. Furthermore, we relate these derivative interactions to
derivative terms in a recently proposed
finite $T$ effective action based on
the SU(3) Wilson Line and its trace, the Polyakov Loop. By this procedure we thus obtain
a perturbative matching at finite $T$ between QCD and the effective model. In particular, we
calculate the leading perturbative QCD-correction to the kinetic term for the Polyakov Loop.  
\end{abstract}

\section{Introduction}
\label{I}
At high temperatures QCD is expected to be found in a new phase, the quark-gluon plasma. 
While the thermal excitations are hadrons and glueballs at low $T$,  
the degrees of freedom in the plasma phase are the quarks and gluons. 
This new state of matter is believed to have existed during the first
microseconds after the Big Bang, and much of the recent interest stems
from the fact such conditions may be produced in heavy-ion collisions. Already the 
results from CERN-SPS seem to hint in that direction, and the experiments at higher energies 
at BNL-RHIC have provided a wealth of new interesting results after its first year of running \cite{rhicwebpage}. 
To understand and interpret the experimental signatures in terms of the
evolution of the initial stage after a heavy-ion collision is clearly a very challenging theoretical task.

The most convincing theoretical results that a drastic change in the degrees 
of freedom takes place at a certain $T$ come from lattice studies. In the pure glue theory, 
there is a phase transition between the confined and deconfined phases at a critical 
temperature $T_c \simeq 270$ MeV \cite{latticesu3}. When massless quarks are added, there      
is similarly a phase transition to a chirally symmetric phase at $T_c \simeq 155-175$ MeV \cite{chiral}, where the precise 
value depends on the number of flavors. Recent lattice simulations also suggest that the chiral 
transition is simultaneous with the deconfining one \cite{digal}. At physical quark masses the situation is not
completely clear, in the sense that there may not be a true phase transition but
only a rapid cross-over \cite{transcross}. Nevertheless, lattice simulations have shown that the pressure,
divided by the ideal gas result and plotted against $T/T_c$, is almost independent of the number of flavors
\cite{indepofnf}.

Due to asymptotic freedom, the quark-gluon plasma behaves as an ideal gas at asymptotically high 
temperatures. Up to corrections of the order of 20 percent, this behavior holds down
to temperatures $T\simeq 3T_c$, even though each higher order
term in a straightforward perturbative expansion gives widely different contributions 
in this temperature regime \cite{pertpressure,nieto}. Instead, at $T\leq 5T_c$
one needs a resummed effective theory in terms of quasi particles, the HTL effective action \cite{htl}. 
Such an effective description 
correctly reproduces thermodynamic quantities like the pressure, as measured by the lattice, down to 
approximately $2T_c$ \cite{blaizotetal,peshier}.
    
Despite this progress, it is of course highly desirable to actually have an analytical description at $T\simeq T_c$, close
to the critical temperature. Since the QCD coupling constant $g
\simeq 2.5$ at $T_c$ (using a renormalization scale $\mu =2\pi T$), 
one is presumably forced to consider effective models that go
beyond the fundamental QCD Lagrangian. 
In a recent paper \cite{robmodel}, such an effective theory was constructed in terms of the thermal Wilson Line ${\bf L}$,
\begin{eqnarray}
{\bf L} = {\cal P} \exp \left [ ig\int_0^{\beta} d\tau A_0 (\vec{x}, \tau ) \right ] , \label{wilsonline}
\end{eqnarray}
where ${\cal P}$ denotes path ordering, $\beta$ is the inverse temperature and
$A_0 = A_0^a T^a$ the time component of the gluon field, with $T^a$ the generators of 
the fundamental SU(3) representation, $a=1,\ldots ,8$. The trace of the Wilson Line is proportional to the 
Polyakov Loop $l$, $l=(1/3){\rm Tr}\,{\bf L}$. 
In the pure Yang-Mills theory, the Polyakov Loop is an order parameter for a global Z(3) symmetry separating the 
confined and deconfined phases, with $\langle l\rangle \neq 0$ ($\langle l\rangle =0$) above (below)
the phase transition \cite{refsforl}. When dynamical quarks are introduced, $l$ ceases to be an order
parameter in the strict sense, but the susceptibility of $l$ still peaks strongly at $T_c$ \cite{digal,lwithquarks}.

In the effective theory \cite{robmodel}, the pressure of the quark-gluon plasma at $T>T_c$ is 
completely due to the condensate of 
$l$, and below $T_c$, where $\langle l\rangle =0$, the pressure vanishes. Moreover, the effective 
potential $V(l)$ changes extremely rapidly around $T_c$. Hence, as the system cools it may find itself 
trapped at the wrong value of $\langle l\rangle$. By coupling the effective field $l$ to hadronic degrees of 
freedom, e.g. the pions, hadrons can be produced as $l$ evolves 
from $\langle l\rangle \neq 0$ and subsequently oscillates around $\langle l\rangle =0$. 
This scenario is somewhat reminiscent of reheating after inflation \cite{linde}, and much
attention has lately been paid to that aspect of the model \cite{robadrian,particleprod}.
Remarkably, many qualitative features observed at RHIC are in accordance with the model predictions.

However, when it comes to questions related to the change of the expectation value of
$l$ and particle production around $T_c$, one has to take into account the variation of
$l$ in space-time. In this paper we address the question of radiative 
corrections to the spatial variation, by considering the leading one-loop QCD contribution to 
the spatial derivatives of $l$. Previous work \cite{robadrian,particleprod} 
took into account only the classical kinetic term in the (Euclidean) effective action, 
$\Gamma (l) = (1/2)\{ |\del_t l|^2 + |\del_i l|^2\} + V(l)$. While such an approach 
certainly is justified at these preliminary stages, it is important to estimate how much the radiative QCD-effects
can affect $\Gamma (l)$ around $T_c$, where the QCD coupling constant becomes large. 
As for the parameters in the potential
$V(l)$, they can be fitted by comparing to QCD lattice results and so are well defined at all $T$.
The kinetic term, on the other hand, has to be matched to perturbatively calculated terms in QCD, 
and could therefore receive large radiative corrections.    
The magnitude of the first one-loop QCD correction can then hopefully serve as a guideline to the importance
of loop effects, and indicate how reliable the above form of $\Gamma (l)$ is at $T_c$. We want to stress that the 
radiative corrections to be discussed come from QCD, and not from fluctuations in $\Gamma (l)$.

To study the correction to the kinetic term $|\del_i l|^2$, we first consider 
the one-loop induced quartic terms in QCD, that contain four powers of the external field $A_0$ 
and two powers of the external momenta. These terms
contribute to the high $T$, dimensionally reduced QCD effective action $\Gamma (A_0)$ \cite{nadkarni,landsman}, 
and apart from providing a correction to $\Gamma (A_0)$ they can also be related 
to the kinetic term in $\Gamma (l)$. As a byproduct we obtain some additional derivative 
interactions in $\Gamma (l)$. 

The paper is organized as follows. In the next section, we give the perturbative QCD 
calculation that corresponds to the leading derivative interactions in $\Gamma (A_0)$.
In Sec.\ III we make the actual matching from an effective theory in terms of
$A_0$ to the one in $l$, and discuss the validity of the results.
We end with our conclusions and an outlook. Our conventions 
and some technical details can be found in the appendix.

\section{Perturbative calculation of the four-point function}
\label{II}
 
In the high temperature regime, long distance phenomena (i.e. $|\vec{x}|\gg\beta$) are dominated by the static
sector of QCD. At high $T$ it therefore makes sense to use dimensional reduction and 
integrate out all the nonstatic modes in the theory \cite{dimred}. 
With only the static modes left, the full QCD Lagrangian is reduced to a three-dimensional theory. In principle
the integrating-out procedure gives rise to an infinite number of interaction terms, but higher dimensional
operators become more suppressed by powers of the QCD coupling constant $g$ and/or
$T$. In full QCD, the following terms in the resulting effective
action $\Gamma (A_0)$ have been calculated \cite{nadkarni,landsman},
\begin{eqnarray}
\Gamma (A_0) = \beta \!\int \! d^3x \left [\frac{1}{2}{\rm Tr}\,F_{ij}^2 + {\rm Tr}\, [D_i,A_0][D_i,A_0] 
+ g^2T^2\left ( 1+\frac{N_{\! f}}{6}\right ){\rm Tr}\, A_0^2 + 
\frac{g^4(9-N_{\! f})}{24\pi^2} \left ( {\rm Tr}\, A_0^2\right )^2 \right ] \ , \label{knownaction}
\end{eqnarray}
where $i,j=1,2,3$, $F_{ij}= F_{ij}^aT^a = (\del_iA_j^a-\del_jA_i^a -gf^{abc}A_i^bA_j^c)T^a$, 
$D_i = \del_i +igA_i$ and $A_i=A_i^aT^a$.
The next term in $\Gamma (A_0)$ contains two derivatives and four powers of
$A_0$, and corresponds to the following part in the Euclidean effective action for $A_0(\vec{x})$,
\begin{eqnarray}
\Gamma^{(4)}_{\!E} (A_0)= \frac{\beta}{24} \prod_{i=1}^4\int \!\frac{d^3k_i}{(2\pi )^3} \, \delta^{(3)}\!(k_1+\ldots +k_4)
\left [ -i\Gamma^{abcd}_{0000} (\vec{k_1},\ldots ,\vec{k_4})\right ]
A_0^a(\vec{k_1})A_0^b(\vec{k_2})A_0^c(\vec{k_3})A_0^d(\vec{k_4}) \ , \label{effactiona0}
\end{eqnarray}
where $\Gamma^{abcd}_{0000}$ is the four-point function of order $O(\beta^2k^2)$, obtained by integrating out all the
non-static modes. Such higher dimensional terms have been calculated in the
pure Yang-Mills theory using the background field method \cite{chapman},
as well as in QED \cite{landsman}, but not in QCD with quarks. 
In this paper we will use a diagrammatic approach to the four-point function.

Perturbatively, the four-point function receives contributions from the diagrams shown
in Fig. 1, together with the additional permutations of the external legs.
There are five permutations adding to the graphs (a), (b) and (d), 
and two to the diagrams (c) and (e), where (e) has a symmetry factor $1/2$.     
Even though all diagrams are superficially logarithmically divergent, it is well known that 
both the fermion diagram and the sum of the pure Yang-Mills diagrams are ultra-violet finite. We will therefore 
only give explicit results for the finite $T$ part of these diagrams, where we  
use the imaginary time formalism \cite{finitetbooks} 
combined with the particular technique described in the appendix.

\begin{figure}[h]
\epsfxsize = 8 cm
\hspace{1cm}
\epsfbox{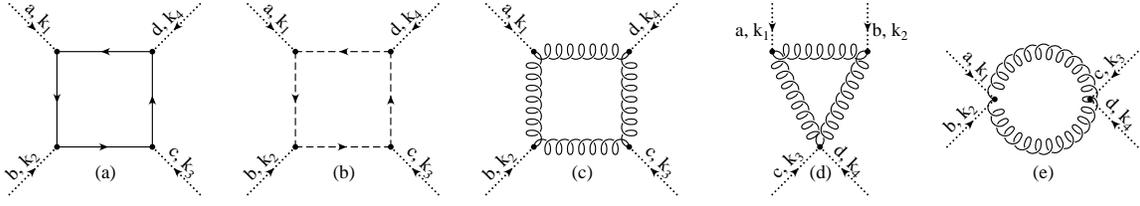}
\vspace{0.1cm}
\caption{{\em Contributions to $\Gamma^{abcd}_{0000}$: 
{\rm (a)} fermions, {\rm (b)} ghosts and {\rm (c)}, {\rm (d)}, {\rm (e)} gluons.}}
\end{figure}

\subsection{The fermion contribution}

Consider first the contribution from the $N_{\! f}$ massless quarks,
with the particular ordering of momenta shown in Fig.\ 1(a). 
Using the conservation of momentum, $k_4 = -(k_1+k_2+k_3)$, with $k_i = \vec{k}_i$, we find
\begin{eqnarray}
\left . \Gamma^{abcd}_{0000} \right |_{(a)_1} = &&\left ( \frac{3ig^4N_{\! f}}{4\pi^{5/2}} \right ) {\rm Tr}_c (T^aT^bT^cT^d)
\int_0^1\!\!dy_1\!\int_0^{1-y_1}\!\!\!dy_2\!\int_0^{1-y_1-y_2}\!\!\!\!dy_3\! \int_{\epsilon-i\infty}^{\epsilon+i\infty}\!\!
\frac{dz}{2\pi i} (1-2^{1-z})\Gamma (z)\xi (z) \cos [\pi z/2]\beta^{-z} \times \nonumber \\
&& \left [ f_1^{-z/2}\Gamma (z/2) \left \{ \frac{5}{2}\Gamma ((1\!-\!z)/2) -6\Gamma ((3\!-\!z)/2) +\frac{2}{3}
\Gamma ((5\!-\!z)/2) \right \} +f_1^{-(2+z)/2}\Gamma ((2\!+\!z)/2) \times \right . \nonumber \\
&& \left . \left \{ f_3\Gamma ((1\!-\!z)/2)-\frac{2}{3}f_2\Gamma ((3\!-\!z)/2)\right \} +\frac{2}{3}f_4f_1^{-(4+z)/2}
\Gamma ((1\!-\!z)/2)\Gamma ((4\!+\!z)/2) \right ] \ , \label{fermionloop}
\end{eqnarray}
where $\xi (z)$ is the Riemann Zeta-function and $f_i = f_i(y_1,y_2,y_3,\vec{k}_1,\ldots ,\vec{k}_3)$, 
given explicitly in the appendix,
are functions of the Feynman parameters $y_k$ and the external momenta $\vec{k}_j$. In the high temperature
limit, where $|\vec{k}_i|\ll T$, we can use the residue theorem to evaluate the $z$-integral by closing the contour
on the left side in the complex $z$-plane. 
Although there is seemingly a logarithmic dependence on $T$ from a double-pole at $z=0$, coming
from the product $\Gamma (z)\Gamma (z/2)$, the coefficient is actually proportional to $[(5/2)\Gamma (1/2) 
-6\Gamma (3/2) +(2/3)\Gamma (5/2)] = 0$. This is in accordance with the fact that the fermion loop does not have any 
logarithmic enhancements \cite{nologs}. 

For the term of order $\beta^2$ we get, from the poles at $z=-2$,
\begin{eqnarray}
\left . \Gamma^{abcd}_{0000} \right |_{(a)_1} = \left ( \frac{-7ig^4N_{\! f}\xi (3)\beta^2}{96\pi^4} \right )
{\rm Tr}_c (T^aT^bT^cT^d) \left ( k_1^2+2k_2^2+k_3^2 +2k_1k_2+2k_2k_3 \right ) \ .
\end{eqnarray}
When the additional five permutations are added and the resulting four-point function inserted into 
Eq.\ (\ref{effactiona0}), the contribution to the effective action from the quark loop becomes,    
\begin{eqnarray}
\left . \Gamma_{\!E} \right |_{(a)} (A_0)= \frac{7g^4N_{\! f}\xi (3)\beta^2}{2304\pi^4}\int_0^{\beta}\!d\tau\!\int \! d^3x \! 
\left [ 2A_0^aA_0^a(\del_iA_0^b)^2 +(\del_i(A_0^aA_0^a))^2 
-2f^{acm}f^{bdm} (\del_iA_0^a)\!\cdot \!(\del_iA_0^b) A_0^cA_0^d \right ] \ , \label{fermionpart1}
\end{eqnarray}
after a partial integration.

In the QED case we have $N_{\! f}{\rm Tr}_c (T^aT^bT^cT^d) \rightarrow 1$, and then 
our result for the two-derivative part of the QED effective action agrees 
with the earlier calculation in \cite{landsman}.

\subsection{The pure Yang-Mills contribution}

We now turn to the pure Yang-Mills contribution, i.e. the diagrams (b)-(e) in Fig. 1. To evaluate the finite $T$
part of these diagrams we will use the gauge condition $\del \! \cdot \! A^a = 0$, and work in Feynman gauge.
As in the fermion case, the functions $g_i = g_i(y_1,\ldots ,\vec{k}_1,\ldots ,\vec{k}_3)$ 
below are all functions of the relevant Feynman parameters and the external momenta. Their 
explicit forms can also be found in the appendix. 
Proceeding in a way similar to the previous section, we have for the ghost loop depicted in Fig.\ 1(b),
\begin{eqnarray}
\left . \Gamma^{abcd}_{0000} \right |_{(b)_1} = && \frac{-ig^4}{8\pi^{5/2}}(f\!f\!f\!f) \!
\int_0^1\!\!dy_1\!\int_0^{1-y_1}\!\!\!dy_2\!\int_0^{1-y_1-y_2}\!\!\!\!dy_3\! 
\int_{1+\epsilon-i\infty}^{1+\epsilon+i\infty}\!\!
\frac{dz}{2\pi i} \Gamma (z)\xi (z) \cos [\pi z/2](\beta^2g_1)^{-z/2} \Gamma ((5\!-\!z)/2)\Gamma (z/2) , \label{ym1}
\end{eqnarray}
where the color structure is $f\!f\!f\!f = f^{fae}\!f^{ebg}\!f^{gch}\!f^{hdf} = \delta_{ab}\delta_{cd}\!
+\! \delta_{ad}\delta_{bc} \!+\!N(d_{abm}d_{cdm} \!-\!d_{acm}d_{bdm}\!+\!d_{adm}d_{bcm})/4$, with $N=3$ and $d_{ijk}$ 
the completely symmetric structure constant. 

For the graph in (c) we find,
\begin{eqnarray}
\left . \Gamma^{abcd}_{0000} \right |_{(c)_1} = &&\left ( \frac{3ig^4}{16\pi^{5/2}} \right ) (f\!f\!f\!f)
\int_0^1\!\!dy_1\!\int_0^{1-y_1}\!\!\!dy_2\!\int_0^{1-y_1-y_2}\!\!\!\!dy_3\! 
\int_{1+\epsilon-i\infty}^{1+\epsilon+i\infty}\!\!
\frac{dz}{2\pi i} \Gamma (z)\xi (z) \cos [\pi z/2]\beta^{-z} \times \nonumber \\
&& \left [ g_2^{-z/2}\Gamma (z/2) \left \{ 5\Gamma ((1\!-\!z)/2) +16\Gamma ((3\!-\!z)/2) +32
\Gamma ((5\!-\!z)/2) \right \} -g_2^{-(2+z)/2}\Gamma ((2\!+\!z)/2) \times \right . \nonumber \\
&& \left . \left \{ g_4\Gamma ((1\!-\!z)/2)-\frac{16}{3}g_3\Gamma ((3\!-\!z)/2)\right \} -\frac{2}{3}g_5g_2^{-(4+z)/2}
\Gamma ((1\!-\!z)/2)\Gamma ((4\!+\!z)/2) \right ] \ , \label{ym2}
\end{eqnarray}
where the color structure is as for the ghost loop.

For the triangle diagram (d) we get,
\begin{eqnarray}
\left . \Gamma^{abcd}_{0000} \right |_{(d)_1} = &&\left ( \frac{-ig^4}{8\pi^{5/2}} \right )(f\!f\!)(f\!f\!+\!f\!f)
\int_0^1\!\!dy_1\!\int_0^{1-y_1}\!\!\!dy_2\!
\int_{1+\epsilon-i\infty}^{1+\epsilon+i\infty}\!\!
\frac{dz}{2\pi i} \Gamma (z)\xi (z) \cos [\pi z/2](\beta^2g_6)^{-z/2}
\times \nonumber \\ && \left [g_6^{-1}(g_6-3g_7)
\Gamma ((1\!-\!z)/2)\Gamma ((2\!+\!z)/2) -15\Gamma ((3\!-\!z)/2)\Gamma (z/2) \right ] \ , \label{ym3}
\end{eqnarray}
where $(f\!f\!)(f\!f\!+\!f\!f) = f^{hfc}\!f^{gdf}(f^{age}\!f^{bhe}\!+\!f^{ahm}\!f^{bgm}) = -2\delta_{ab}\delta_{cd}\!
-\!\delta_{ac}\delta_{bd} \!-\!\delta_{ad}\delta_{bc}\!-\!Nd_{abm}d_{cdm}/2$. 

Finally, we find for graph (e),
\begin{eqnarray}
\left . \Gamma^{abcd}_{0000} \right |_{(e)_1}= &&\left ( \frac{3ig^4}{16\pi^{5/2}} \right )
(f\!f\!+\!f\!f)(f\!f\!+\!f\!f)\int_0^1\!\!dy_1\!\int_{1+\epsilon-i\infty}^{1+\epsilon+i\infty}\!\!
\frac{dz}{2\pi i} \Gamma (z)\xi (z) \cos [\pi z/2](\beta^2g_8)^{-z/2}\Gamma ((1\!-\!z)/2)\Gamma (z/2) \ , \label{ym4}
\end{eqnarray}
with $(f\!f\!+\!f\!f)(f\!f\!+\!f\!f) = (f^{age}\!f^{bhe}\!+\!f^{ahe}\!f^{bge})
(f^{cgf}\!f^{dhf}\!+\!f^{chf}\!f^{dgf}) = 4\delta_{ab}\delta_{cd}\!+\!2\delta_{ac}\delta_{bd} \!+\!2
\delta_{ad}\delta_{bc}\!+\!Nd_{abm}d_{cdm}$. 

Contrary to the fermion case, at order $O(\beta^0)$ each diagram contains a logarithmic dependence 
on $T$ and the external momenta $\vec{k}_j$. 
In addition, the $T=0$ part depends logarithmically on $\vec{k}_j$
and an ultraviolet cut-off $\Lambda$, that has to be introduced to regularize the loop-momentum integral. 
The logarithmic dependence on $\vec{k}_j$ cancels out between the $T=0$ and $T>0$ parts for each 
permutation of each individual diagram, whereas the terms containing $\log T$ ($\log \Lambda $) only cancel out in
the total $T>0$ ($T=0$) result, i.e. when all the different diagrams are added together. This was basically noted
already in \cite{nadkarni}, and we have checked that it holds true in our calculations as well. 
There is also a linear divergence when $\beta k\rightarrow 0$, at order $O(1/\beta)$, 
in all of the Eqs.\ (\ref{ym1})-(\ref{ym4}). This divergence originates
from static propagators running in the loop, i.e. the propagators with a vanishing Matsubara frequency, $\omega_n=0$. 
If only the non-static modes are integrated out, the static terms should be subtracted and our remaining result is
then finite in the limit $\beta k\rightarrow 0$, as it should \cite{nadkarni}. 
We emphasize that the $\omega_n=0$ modes do not influence the two-derivative term.
 
Taking into account all the permutations of Eqs.\ (\ref{ym1})-(\ref{ym4}) and adding the different 
contributions, we find for the $O(\beta^2)$ term,
\begin{eqnarray}
\left . \Gamma_{\!E} \right |_{(b)-(e)}(A_0) = -\frac{g^4\xi (3)\beta^2}{256\pi^4}\int_0^{\beta}\!d\tau\!\int \! d^3x \! 
\left [ 2A_0^aA_0^a(\del_iA_0^b)^2 +(\del_i(A_0^aA_0^a))^2 +\frac{11}{6}
f^{acm}f^{bdm} (\del_iA_0^a)\!\cdot \!(\del_iA_0^b) A_0^cA_0^d \right ] \ , \label{gluepart}
\end{eqnarray}
where we have performed an integration by parts. This 
result for the pure Yang-Mills contribution disagrees slightly with the
previous finding from the background field method
in \cite{chapman}, in that the first two term in Eq.\ (\ref{gluepart}) are a factor
$(2/11)$ smaller, and the last a factor $(-1/2)$\footnote{It should be noted, 
however, that the discrepancy is of marginal practical importance when it comes to the qualitative
discussion of the influence of these QCD-terms in the Polyakov Loop action.}.
\subsection{The total contribution}

By combining the results in Eqs.\ (\ref{fermionpart1}) and (\ref{gluepart}), the complete 
contribution to the effective action becomes,
\begin{eqnarray}
\Gamma_{\!E}^{(4)} (A_0)= && \left . \Gamma_{\!E} \right |_{(a)} (A_0)+ \left . \Gamma_{\!E} \right |_{(b)-(e)} (A_0)= 
\frac{g^4\xi (3)\beta^2}{256\pi^4}\left ( \frac{7N_{\! f}}{9}-1\right )\int_0^{\beta}\!d\tau\!\int \! d^3x \! 
\left [ 2A_0^aA_0^a(\del_iA_0^b)^2 +(\del_i(A_0^aA_0^a))^2\right ] - \nonumber \\
&& \frac{11g^4\xi (3)\beta^2}{1536\pi^4}\left ( \frac{28N_{\! f}}{33}+1\right )\int_0^{\beta}\!d\tau\!\int \! d^3x \!
\left [ f^{acm}f^{bdm} (\del_iA_0^a)\!\cdot \!(\del_iA_0^b) A_0^cA_0^d \right ] \ . \label{effaction}
\end{eqnarray}
In the dimensionally reduced theory of QCD \cite{nadkarni,landsman}, Eq.\ (\ref{effaction}) provides the leading 
derivative interactions between the $A_0$-fields. How large the calculated derivative term is, compared to 
both the ones already present in Eq.\ (\ref{knownaction}) as well as the omitted
higher dimension operators, depends on the scales of interest. For instance, at the soft scale where $\del_i 
\sim gT$ and $A_0 \sim T$, we have $g^4\beta^2\del^2A_0^4 \sim g^6T^4$, a factor $g^2$ higher than the term
$g^4A_0^4$.
 
In general, this effective theory is interesting on length scales 
$|\vec{x}|\ll \beta$ in the high temperature regime, especially when combined with nonperturbative lattice 
methods \cite{kajantie}. It is for example possible to study the non-perturbative Debye mass 
\cite{debyemass}, and the 3d effective theory is also useful for 
calculations of the pressure in the quark-gluon phase, both perturbatively \cite{nieto} and nonperturbatively
\cite{resum}. Although the derived correction in Eq.\ (\ref{effaction}) presumably gives a minor 
effect only, it is somewhat interesting to note that the terms are rather sensitive to the number of quark flavors. 
In Eq.\ (\ref{effaction}), the coefficient of the first term is proportional to $(7N_{\! f} -9)$, which 
goes from $-1$ to $4/3$ between $N_{\! f}=0$ and $N_{\! f}=3$. Similarly, the second term in
Eq.\ (\ref{effaction}) increases by more than 250\% in the same range of $N_{\! f}$. 
This should be contrasted with the constant part of the $A_0^4$-contribution to 
Eq.\ (\ref{knownaction}), that depends on $N_{\! f}$ as $(9 - N_{\! f})$ and therefore
only changes by 33\% when going from $N_{\! f}=0$ to $N_{\! f}=3$.
Due to this strong behavior of the number of quark flavors, it is not
inconceivable that the derivative interactions will make a small but noticeable difference between e.g. the pure
glue theory and the three-flavor case.

\section{Derivative terms in the Wilson Line model}

The QCD dimensionally reduced theory describes accurately static phenomena at very high $T$, but
the approximations break down around a few times $T_c$ \cite{kajantie}. 
In addition, one is by construction omitting all dynamical information.

To understand the features around $T_c$ a Ginzburg-Landau type of effective theory was 
proposed in \cite{robmodel}. In this model, the potential is written in terms of $l$,
\begin{eqnarray}
V(l)= a_1T^4\left [ -a_2|l|^2 -a_3\left ( l^3+{\rm c.c.} \right )+|l|^4 
\right ] \ . \label{robpotential}
\end{eqnarray}
The constants $a_i$ are then used to fit the pressure above $T_c$, with $a_2$ a function of 
temperature so that the global minimum of the potential is at $l\neq 0$ ($l=0$) above (below) $T_c$ \cite{robadrian}.
One of the important aspects of the potential in Eq.\ (\ref{robpotential}) is the extremely rapid change around
$T_c$, due to a very sensitive dependence of $a_2$ on $T/T_c$ \cite{robadrian}. In a dynamical scenario one can therefore
assume an instantaneous quench, where the value of $l$ suddenly no longer corresponds to the correct minimum.
The $l$-field then rolls down the potential, and by coupling the $l$-field to a linear
sigma model the potential energy is converted into pions \cite{robadrian,particleprod}. Even though the model 
is of phenomenological origin, it thus makes predictions that can be compared to experimental results.

After the quench, the evolution of the Euler-Lagrange equations from the initial conditions requires, 
apart from the potential and the coupling to the 
chiral field, also a kinetic term for $l$ \cite{robadrian,particleprod}. Although the time dependence
is beyond the calculation presented in this paper, we can provide the first 
perturbative QCD-correction to the spatial derivatives. The leading coefficient is
the classical contribution to the derivative term, and comes from the kinetic term 
of $A_0 (\vec{x})$ in Eq.\ (\ref{knownaction}), as can be seen from the following argument \cite{robprivate}: 
decomposing the Wilson Line in Eq.\ (\ref{wilsonline}) into an octet $\tilde{{\bf L}}$ 
and the singlet $l$,
\begin{eqnarray}
{\bf L} = \tilde{{\bf L}} + {\bf 1}\frac{1}{N}{\rm Tr}\, {\bf L} = \tilde{{\bf L}} + {\bf 1}l \ ,
\end{eqnarray}
where $\tilde{{\bf L}}$ is traceless and $N=3$, we have
\begin{eqnarray}
{\rm Tr}|\del_i {\bf L}|^2 = {\rm Tr}|\del_i \tilde{{\bf L}}|^2 + 3|\del_i l|^2 \ . \label{decomposeL}
\end{eqnarray}
On the other hand, by a direct calculation in the static limit,
\begin{eqnarray}
{\rm Tr}|\del_i {\bf L}|^2 = g^2\beta^2 (\del_iA_0^a)^2 + {\rm Tr}\, \left \{ {\rm commutator \, terms} 
\right \} \ . \label{commutators}
\end{eqnarray}
When the commutator terms in Eq.\ (\ref{commutators}) are rewritten in terms of ${\bf L}$ they can only 
involve the adjoint field, or products of $l$ and $\tilde{{\bf L}}$, since $l$ (times the identity
matrix) by itself commutes 
with all SU(3) matrices. Thus, by combining Eqs.\ (\ref{decomposeL}) and (\ref{commutators}), we have,
\begin{eqnarray}
(1/2)(\del_iA_0)^2 = \frac{T^2}{2g^2}\left [ {\rm Tr}|\del_i \tilde{{\bf L}}|^2 + 3|\del_i l|^2 \right ] +
f(\tilde{{\bf L}}) \ , \label{leadingderivative}
\end{eqnarray}
where $f(\tilde{{\bf L}})$ corresponds to the commutator terms, rewritten as a function of $\tilde{{\bf L}}$.
At $T_c$, $g\simeq 2.5$ so that the leading coefficient for the kinetic term of $l$ is $3/2g^2 \simeq 0.3$, 
which is reasonably close to the canonical value $1/2$. Given the unknown function $f(\tilde{{\bf L}})$ 
it is not clear whether the kinetic term for the adjoint field actually is unique. However, $\tilde{{\bf L}}$ does
not play any important role around $T_c$, and can therefore be neglected on physical grounds \cite{robmodel}. 

The procedure to obtain the kinetic term for $l$ is thus to match terms in the effective theory $\Gamma (A_0)$
to a corresponding $|\del_i l|^2$ piece in $\Gamma (l)$. This means that
the classical coefficient for $|\del_i l|^2$ will change when radiative corrections are taken into
account in $\Gamma (A_0)$. To lowest order, the kinetic term for $A_0$ can receive corrections from the polarization 
tensor \cite{nadkarni,landsman}, that would affect the $|\del_i l|^2$ term via Eq.\ (\ref{leadingderivative}). However,
with the optimal choice for the counterterms \cite{landsman} there is in fact no $O(g^2)$-contribution, so the 
renormalized kinetic term in $\Gamma (A_0)$ remains $(1/2)(\del_i A_0)^2$. 

Even though the kinetic term for $A_0$ is unchanged at one-loop, this does not mean that $|\del_i l|^2$ is so. 
Since $l = (1/3){\rm Tr}\,{\bf L}$, the  
$|\del_i l|^2$-term contains at least four powers of $A_0$ when ${\bf L}$ is expanded in powers of $A_0$. 
The two-derivative term in $\Gamma (l)$ can therefore receive a perturbative correction 
from the four-point function in Eq.\ (\ref{effactiona0}). Indeed, by using the relations
\begin{eqnarray}
A_0^aA_0^a(\del_iA_0^b)^2 = &&-\frac{12}{g^4\beta^4}\left [ |l|^2 -1\right ] \left (
{\rm Tr} |\del_i \tilde{{\bf L}}|^2 +3|\del_i l|^2 \right ) +O(A_0^6) \ , \nonumber \\ 
 (\del_i(A_0^aA_0^a))^2 = && \frac{144}{g^4\beta^4}|\del_i l|^2 +O(A_0^6)\nonumber \\
 f^{acm}f^{bdm} (\del_iA_0^a)\!\cdot \!(\del_iA_0^b) A_0^cA_0^d = && 2{\rm Tr}|
[\del_i \tilde{{\bf L}}, \ \tilde{{\bf L}}] |^2 +O(A_0^6) \ ,
\end{eqnarray}
we can rewrite Eq.\ (\ref{effaction}) as 
\begin{eqnarray} 
\Gamma_{\!E}  = && \int_0^{\beta}\!d\tau\!\int \! d^3x \! \left \{ \frac{\xi (3)T^2}{16\pi^4}\left 
(\frac{7N_{\! f}}{9}-1\right )
\left [ \frac{3}{2} (1-|l|^2){\rm Tr}|\del_i\tilde{{\bf L}}|^2 +\frac{27}{2}|\del_i l|^2
-\frac{9}{2}|l|^2 |\del_i l|^2 \right ] \right . - \nonumber \\ && \left .
\frac{11\xi (3)T^2}{768\pi^4} \left (\frac{28N_{\! f}}{33}+1\right )
{\rm Tr}|[\del_i \tilde{{\bf L}}, \ \tilde{{\bf L}} ]|^2 \right \} \ . \label{gammainls}
\end{eqnarray}  
From the calculation of the four-point function we thus find the leading perturbative correction to the kinetic term
of $l$. Including the contribution from Eq.\ (\ref{gammainls}), the kinetic term now becomes 
\begin{eqnarray}
\frac{3T^2}{2g^2}|\del_i l|^2 \rightarrow 
\frac{3T^2}{2g^2}\left [ 1+ \frac{g^2\xi (3)}{16\pi^4}(7N_{\! f}-9)\right ] |\del_i l|^2 =
\frac{3T^2}{2g^2} \left [ 1+c_{{\rm corr}}\right ] |\del_i l|^2 \ .
\end{eqnarray}
Since $l$ is dimensionless, the correct dimension of the operator has to be supplied by some 
other scales. In the perturbative calculation the only scale is $T$, and hence the one-loop induced
coefficient has the same $T$-dependence as the leading term in Eq.\ (\ref{leadingderivative}).
However, the perturbative correction does not depend on the QCD coupling constant and is therefore just 
a fixed number at this order. For example, for three flavors the coefficient is $\simeq 1.4\times 10^{-2}T^2$.
Compared to the classical contribution $(3/2g^2)T^2$, the fraction of 
the one-loop correction is only $0.01g^2$. 
Even at $T_c$, this is merely of the order of 5\%, and at higher $T$ even less due to the logarithmic decrease of $g$.    

Having derived the first correction to the kinetic term for $l$ from perturbation theory,
let us now discuss to what extent, and in what temperature range, the terms in Eq.\ (\ref{gammainls})
can be trusted. First of all, it should be noted that the form of the effective action in Eq.\ (\ref{gammainls}) is not 
completely unique. The reason is that by a partial
integration, and discarding any surface terms, we can always trade factors of $[A_0^aA_0^a(\del_iA_0^b)^2]$
and $(\del_i(A_0^aA_0^a))^2$ for a term like
$[A_0^aA_0^aA_0^b(\del_i^2\!A_0^b)]$, but this equality does not hold at the level of $l$. In fact,
if a term $[A_0^aA_0^aA_0^b(\del_i^2\!A_0^b)]$ is kept in the action, not only do the 
coefficients in Eq.\ (\ref{gammainls}) change, but there are also additional nonequivalent terms of the form
$(\del_i|l|^2)^2$ and $|l|^2[\del_i^2(l+l^{\ast})]$. 
Nevertheless, the action in Eq.\ (\ref{gammainls}) is of course unique to order $O(A_0^4)$.
Since higher order operators in the dimensionally reduced theory are further suppressed at high $T$,
the predictions in Eq.\ (\ref{gammainls}) should at the very least be reliable down to $T\geq 2T_c-3T_c$, i.e. when
the effective 3d theory itself is applicable.

When $T\rightarrow T_c$, the question is admittedly more subtle, as higher loop effects, higher dimensional operators
and possibly nonperturbative effects become important. However, Eq.\ (\ref{gammainls}) does not have to
break down completely when the 3d theory does so. The 3d theory becomes invalid because the procedure 
of integrating out the non-static modes is unreliable when $g^2(T)\, T \simeq \pi T$ \cite{kajantie}.
In contrast, $\Gamma (l)$ is by construction valid near $T_c$, so the 
question is rather how much the coefficient for the spatial derivative term changes.

Considering first the operators of higher dimensionality, it is certainly possible to
imagine that their bulk part follows from 
an expansion of Eq.\ (\ref{gammainls}). The additional contributions that do not originate from 
these sources, e.g. terms like $T^2(|l|^2-1)^n|\del_il|^2$, that are at 
least of order $A_0^{(2n+2)}$ (with $n\geq 2$), would then
be suppressed. Not because they are unimportant a priori, but because their numerical coefficients are small. 
There are also higher derivative terms not accounted for in Eq.\ (\ref{gammainls}), like 
$(|l|^2-1)(\del_i^2|l|^2)^2$, but they do not affect the kinetic term. To $O(A_0^4)$, 
they are in fact straightforward to obtain from Eq.\ (\ref{fermionloop}) and Eqs.\ (\ref{ym1})-(\ref{ym4}).

When it comes to higher loop and nonperturbative effects, they will naturally induce a $T$-dependence
in the radiative corrections to the kinetic term. 
Thus, $(3T^2/2g^2)(1+c_{\rm corr})|\del_il|^2\rightarrow (3T^2/2g^2)(1+c_{\rm corr}(T/T_c, T/\Lambda ))|\del_il|^2 $, 
which follows partly from the running of the QCD coupling constant in the higher loop effects. 
How much this will affect the kinetic term is difficult to estimate, but the small correction 
from the four-point function may indicate that the perturbative QCD-contributions are not too important for 
constructing $\Gamma (l)$. 

\section{Summary and Conclusions}
In this paper we calculated the leading momentum dependence of the four-point function in QCD with
$N_{\! f}$ massless flavors, and related this contribution to terms in both the effective action $\Gamma (A_0)$
and $\Gamma (l)$.

As for the derivative terms in $\Gamma (A_0)$, they will only have a minor influence when $T\gg T_c$.
As the temperature decreases and approaches $T_c$ the 3d theory becomes less reliable,
but there could very well be a temperature region where the effective theory is still valid and the derivative interactions 
nonnegligible. Since the contribution has a rather strong dependence on 
$N_{\! f}$, a difference between the pure glue theory and e.g. $N_{\! f}=3$ QCD could perhaps be noticed.

From the four-point function, we also found the lowest one-loop QCD correction to the 
spatial derivative term in the effective theory $\Gamma (l)$. The coefficient is
independent of $g$ and much smaller than the classical term, the ratio between the two being 
of the order of $10^{-2}$ at $T_c$. This derivation assumes that the
strange quark mass $m_s$ can be neglected even at $T_c$, which of course is
an oversimplification, given that $m_s\sim T_c$. Nevertheless,
the influence of $m_s$ is not likely to change the fact that the
correction is small even at $T_c$.  

At one-loop, there is an infinite number of terms that contribute to the coefficient of $|\del_i l|^2$, to 
the same order in $g$ and $T$, as the four-point function. This follows from the fact the
induced two-derivative interactions, with $n$ external fields $A_0$, is of the functional form 
$g^n\del^2 A_0^n/T^{(n-2)}$, which corresponds to an expansion of $l$ to at most order $(n-2)$ in the term 
$T^2|\del_i l|^2$. Some of these higher-dimensional contributions, maybe even the major parts, 
are already accounted for by rewriting the four-point
function in $\Gamma (A_0)$ in terms of $l$, as in Eq.\ (\ref{gammainls}). 
In any case, since the four-point contribution is very small, it is reasonable to assume that 
the higher $n$-point functions give even smaller corrections. In that case, Eq.\ (\ref{gammainls}) should give the
correct order of magnitude for the total one-loop correction. 

As mentioned earlier, there are also higher loop effects that contribute to the kinetic term in $\Gamma (l)$.
For example, taking into account the two-loop correction to the diagrams in Fig.\ 1 gives
$c_{{\rm corr}} \rightarrow c_{{\rm corr}}[1+ag^2]$. To understand the reliability
of the canonical term it is then crucial to know the magnitude of $a$. Surprisingly, studies of 
higher loop effects in the 3d effective theory $\Gamma (A_0)$ indicate that they only give corrections of the order of 
30\% at $T_c$ \cite{kajantie}. If these conjectures can be taken over to the Wilson Line model, one could
in fact expect the derivative term $(1/2)|\del_i l|^2$ to change by perhaps at most a factor two,
with all QCD-corrections taken into account. Of course, this has to be regarded as a highly speculative 
suggestion at the present stage.    

To complete the dynamical scenario one also needs the time dependence of $l$. 
Unfortunately, it is yet unclear how $l$ generalizes to a real time formulation \cite{robmodel}.
Assuming that the form of the spatial derivatives can be extended to a Lorentz invariant form,  
the predictions for pion production and the evolution of $|l|$ will remain almost
unchanged \cite{robadrian,particleprod}. In particular, if the Lorentz invariant kinetic term does not
change by more than a factor of two, it can easily be compensated by a difference in e.g. the 
expansion rate of the plasma. 

Finally, to obtain a decisive estimate of the QCD-effects in $\Gamma (l)$,
one has to establish either a unique mapping from $\Gamma (A_0)$ to $\Gamma (l)$, or find a way to
determine $\Gamma (l)$ directly, perhaps numerically. Hopefully, the calculations presented in this paper   
can serve as a first step in that direction.

\begin{center}
{\bf Acknowledgments}
\end{center}
\noindent
The author thanks R.\ D.\ Pisarski for discussions that initialized this project, and for reading the manuscript.
D.\ B\"{o}deker, A.\ Dumitru
and R.\ D.\ Pisarski are greatfully acknowledged for useful and interesting discussions during the 
investigations, and K.\ Kajantie and M.\ Laine for useful comments on
an early version of the manuscript. 
This work was supported by The Swedish Foundation for International Cooperation in Research
and Higher Education (STINT) under contract 99/665, and in part by DOE grant DE-AC02-98CH10886.                   
 
\appendix

\section{Evaluation of the Feynman diagrams}

In this appendix we outline our method for calculating the Feynman diagrams
shown in Fig.\ 1.
We first follow the Feynman rules given in \cite{peskin} for Minkowski 
space-time. After all contractions and traces over spinor indices have been performed, 
we continue to a Euclidean space compact in the imaginary time
direction: 
\begin{eqnarray}
p_0\rightarrow i\omega_n \ , \ \ \ \ \int\frac{dp_0}{2\pi}\rightarrow  
iT\sum_n \ ,  
\end{eqnarray}
where $\omega_n$ is the Matsubara frequency, $\omega_n = (2n+1)\pi T$
($\omega_n=2\pi T$) for fermions (bosons).

Next, we use a Feynman parametrization to combine the denominators in the loop integral, and 
extract the $T>0$ part from the following relations \cite{finitetbooks}:
\begin{eqnarray}
\left . T\sum_n f(p_0=i\omega_n) \right |_{T>0} = \left \{ \matrix{ &  
\int_{\delta-i\infty}^{\delta+i\infty}\frac{dp_0}{2\pi i}
\, n_B \, [f(p_0)+f(-p_0)] \ \ {\rm (bosons)} \cr & \mbox{} \cr
& \int_{\delta-i\infty}^{\delta+i\infty}\frac{dp_0}{2\pi i}
\, n_F \, [f(p_0)+f(-p_0)]
\ \ {\rm (fermions)}} \right . \ .
\end{eqnarray}
where $n_B = [\exp (\beta p_0)-1]^{-1}$, $n_F =[\exp (\beta p_0)+1]^{-1}$ and $\delta \rightarrow 0^{+}$. 
The ghosts follow Bose statistics, despite their anticommuting properties.

By shifting the vector momentum in the loop, $\vec{p}$, to $\vec{q}=\vec{p} + h(y_i, \vec{k}_i)$, with $h(y_i, \vec{k}_i)$
a linear function of the external momenta $\vec{k}_i$,  
we then integrate over $\vec{q}$. Finally, by performing a Mellin transform,
\begin{eqnarray}
\left ( e^x \pm 1 \right )^{-1} = \int_{c_{\pm}-i\infty}^{c_{\pm}+i\infty}\frac{dz}{2\pi i} \Gamma (z)\xi (z)v_{\pm}x^{-z} \ ,
\end{eqnarray}
where $v_{+}=(1-2^{1-z})$, $v_{-}=1$, and with the contour specified by $c_{+}=\epsilon$,
$c_{-}=1+\epsilon$ (where $\epsilon\rightarrow 0^{+}$), we can integrate over $p_0$ after a change of variables.

To illustrate the above procedure, consider the diagram in Fig.\ 1(e). Omitting
the color factors for simplicity, and using the notation $|\vec{p}|=p$,
$|\vec{k}_1+\vec{k}_2|=|\vec{k}_{12}|=k_{12}$, we have to calculate the following integral: 
\begin{eqnarray}
&& \frac{3g^4}{2}\!\int \!\frac{d^4p}{(2\pi )^4}\, \frac{1}{p_0^2-p^2}\, \frac{1}{p_0^2-(p+k_{12})^2} 
\stackrel{T>0}{\rightarrow} 3ig^4 \! 
\int_{\delta-i\infty}^{\delta+i\infty}\! \frac{dp_0}{2\pi i} n_B
\! \int_0^1 \! dy_1\!\int \! \frac{d^3p}{(2\pi )^3}
\frac{1}{[p^2+\{2\vec{p}\cdot \vec{k}_{12} + k_{12}^2 \}(1\!-\!y_1)-p_0^2]^2} 
\nonumber \\ && \ \ \ \ = \frac{3ig^4}{2\pi^2}\int_0^1dy_1
\int_{\delta-i\infty}^{\delta+i\infty}\frac{dp_0}{2\pi i} n_B
\int_0^{\infty}dq \frac{q^2}{[q^2+g_8-p_0^2]^2} \ , \label{illustration}
\end{eqnarray}
where we made the shift $\vec{q}=\vec{p}+(1-y_1)\vec{k}_{12}$ and defined $g_8 = y_1(1-y_1)k_{12}^2$ in the last integral.
After performing 
the $q$-integral in Eq.\ (\ref{illustration}),
we are left with,
\begin{eqnarray}
&& \frac{3ig^4}{8\pi}\! \int_0^1\!dy_1 \int_{\delta-i\infty}^{\delta+i\infty}\frac{dp_0}{2\pi i}\left (
\frac{1}{e^{\beta p_0}\!-\!1}\right )\frac{1}{\sqrt{g_8-p_0^2}} = \frac{3ig^4}{8\pi^2}\!\int_0^1\!dy_1
\int_{1+\epsilon-i\infty}^{1+\epsilon+i\infty}
\frac{dz}{2\pi i}\Gamma (z) \xi (z) \cos [\pi z/2]\beta^{-z} \! \int_0^{\infty}\! du
\frac{u^{-z}}{\sqrt{g_8+u^2}} \nonumber \\ && \ \ \ = \frac{3ig^4}{16\pi^{5/2}} \int_0^1dy_1
\int_{1+\epsilon -i\infty}^{1+\epsilon +i\infty}
\frac{dz}{2\pi i}\Gamma (z) \xi (z) \beta^{-z} \cos [\pi z/2] g_8^{-z/2}
\Gamma ((1-z)/2) \Gamma (z/2) \ ,
\end{eqnarray}
which is Eq.\ (\ref{ym4}), except for the color factors.

To check our method we also calculated the $O(\beta^2k_{12}^2)$ contribution to
the above diagram in a different way: we first performed the $p_0$-integral, by picking up the poles in 
the complex $p_0$-plane, and then did the $p$-integral, 
without any Feynman parametrization, by expanding the integrand differently in different integration regions. 
The two results of course agree with each other.

For completeness, we also give the functions $f_1,\ldots ,f_4$ in Eq.\ (\ref{fermionloop}) and $g_1,\ldots ,g_8$ in Eqs.\ (\ref{ym1})-(\ref{ym4}). Using the conservation of momentum, and the shorthand 
notation $\vec{k}_i+\vec{k}_j+\vec{k}_l=k_{ijl}$, their explicit forms are as follows:
\begin{eqnarray}
f_1 =&& y_2k_1^2+y_3k_{12}^2+(1-y_1-y_2-y_3)k_{123}^2 - [y_2k_1+y_3k_{12}+(1-y_1-y_2-y_3)k_{123}]^2
\\ 
f_2 =&& 3k_1^2+4k_1k_2+k_2^2+2k_1k_3+k_2k_3-3(3k_1+2k_2+k_3)[y_2k_1+y_3k_{12}+(1-y_1-y_2-y_3)k_{123}]+\nonumber \\
&& +6[y_2k_1+y_3k_{12}+(1-y_1-y_2-y_3)k_{123}]^2 \\
f_3 =&& (1/3)\left \{ 5k_1^2y_1(2y_1-1)+k_2^2[3+10y_1^2-10y_2+10y_2^2+10y_1(2y_2-1)]+
+5k_3^2[1+2y_1^2+2y_2^2-3y_3+2y_3^2+\right . \nonumber \\ && +y_2(4y_3-3)+y_1(4y_2+4y_3-3)]+k_2k_3[8+20y_1^2+20y_2^2-10y_3+
5y_2(4y_3-5)+5y_1(8y_2+4y_3-5)] + \nonumber \\ && \left .
k_1[k_2(3+20y_1^2-5y_2+5y_1(4y_2-3))+k_3(2+20y_1^2-5y_2-5y_3+20y_1(y_2+y_3-1))] \right \} \\
f_4 =&& \left \{ [(1-y_2)k_1-y_3k_{12}-(1-y_1-y_2-y_3)k_{123}]\cdot [-y_2k_1+(1-z)k_{12}-(1-y_1-y_2-y_3)k_{123}]\right \}
\times \nonumber \\ &&
\left \{ [-y_2k_1-y_3k_{12}-(1-y_1-y_2-y_3)k_{123}]\cdot [-y_2k_1-y_3k_{12}+(y_1+y_2+y_3)k_{123}]\right \} - \nonumber \\
&& -\left \{ [-y_2k_1-y_3k_{12}-(1-y_1-y_2-y_3)k_{123}]\cdot [-y_2k_1+(1-z)k_{12}-(1-y_1-y_2-y_3)k_{123}] \right \} 
\times \nonumber \\ &&
\left \{ [(1-y_2)k_1-y_3k_{12}-(1-y_1-y_2-y_3)k_{123}]\cdot [-y_2k_1-y_3k_{12}+(y_1+y_2+y_3)k_{123}]\right \}+
\nonumber \\ &&
+\left \{ [-y_2k_1-y_3k_{12}-(1-y_1-y_2-y_3)k_{123}]\cdot [(1-y_2)k_1-y_3k_{12}-(1-y_1-y_2-y_3)k_{123}]\right \} \times
\nonumber \\ &&
\left \{ [-y_2k_1+(1-z)k_{12}-(1-y_1-y_2-y_3)k_{123}]\cdot [-y_2k_1-y_3k_{12}+(y_1+y_2+y_3)k_{123}]\right \} \\
g_1 = && f_1 \\
g_2 = && f_1 \\
g_3 = && 3k_1^2+2k_2^2+2k_3^2+3k_1k_2+3k_1k_3+2k_2k_3+2[(1-y_1)k_1+(1-y_1-y_2)k_2+(1-y_1-y_2-y_3)k_3]^2 - \nonumber \\ &&
- (3k_1+2k_2+k_3)[(1-y_1)k_1+(1-y_1-y_2)k_2+(1-y_1-y_2-y_3)k_3] \\
g_4 = && (1/3)\left \{ k_1^2(10y_1-20y_1^2-7)-2k_2^2[10y_1^2-10y_2+10y_2^2+10y_1(2y_2-1)-3] - \right . \nonumber \\
&& - k_3^2[17+20y_1^2+20y_2^2-30y_3+20y_3^2+10y_2(4y_3-3)+10y_1(4y_2+4y_3-3)] - \nonumber \\
&& -2k_2k_3[2+20y_1^2+20y_2^2-10y_3+
5y_2(4y_3-5)+5y_1(8y_2+4y_3-5)] - \nonumber \\
&& \left . -2k_1[k_2(20y_1^2-5y_2+5y_1(4y_2-3)-3)+5k_3(3+4y_1^2-y_2-y_3+4y_1(y_2+y_3-1))] \right \} \\
g_5 = && \left \{ [y_1k_1+(1+y_1+y_2)k_2-(1-y_1-y_2-y_3)k_3]\cdot [(2-y_1)k_1+(1-y_1-y_2)k_2+(1-y_1-y_2-y_3)k_3] \right \}
\times \nonumber \\ && \left \{ [-y_1k_1-(y_1+y_2)k_2+(2-y_1-y_2-y_3)k_3]\cdot [(2-y_1)k_1+(2-y_1-y_2)k_2
+(2-y_1-y_2-y_3)k_3] \right \} + \nonumber \\ &&
\left \{ [(1+y_1)k_1-(1-y_1-y_2)k_2-(1-y_1-y_2-y_3)k_3]\cdot [(1+y_1)k_1+(1+y_1+y_2)k_2+(1+y_1+y_2+y_3)k_3] \right \}
\times \nonumber \\ &&
\left \{ [-y_1k_1+(2-y_1-y_2)k_2+(1-y_1-y_2-y_3)k_3]\cdot [y_1k_1+(y_1+y_2)k_2+(1+y_1+y_2+y_3)k_3] \right \} \\
g_6 = && y_1k_{12}^2+y_2k_{123}^2+(1-y_1-y_2)k_{1234}^2 -[k_1+k_2+(1-y_1)k_3+(1-y_1-y_2)k_4]^2 \label{eqg6} \\
g_7 = && [(1+y_1)k_3-(1-y_1-y_2)k_4][(2-y_1-y_2)k_4-y_1k_3] \ , \label{eqg7}
\end{eqnarray}
where, in order to simplify the permutations of the triangle graph, we did not use $k_4=-(k_1+k_2+k_3)$ in 
Eqs.\ (\ref{eqg6}) and (\ref{eqg7}). Finally,  
\begin{equation}
\hspace{-13.9cm} \hspace{2pt}
g_8 = y_1(1-y_1)k_{12}^2 \ . 
\end{equation}

\end{document}